\documentclass[12pt]{article}
\usepackage{html}
\usepackage{epsf}
\usepackage{graphicx}
\usepackage{amssymb}
\begin{document}
\title{MATTERS OF GRAVITY, The newsletter of the APS Topical Group on 
Gravitation}
\begin{center}
{ \Large {\bf MATTERS OF GRAVITY}}\\ 
\bigskip
\hrule
\medskip
{The newsletter of the Topical Group on Gravitation of the American Physical 
Society}\\
\medskip
{\bf Number 35 \hfill Winter 2010}
\end{center}
\begin{flushleft}
\tableofcontents
\vfill\eject
\section*{\noindent  Editor\hfill}
David Garfinkle\\
\smallskip
Department of Physics
Oakland University
Rochester, MI 48309\\
Phone: (248) 370-3411\\
Internet: 
\htmladdnormallink{\protect {\tt{garfinkl-at-oakland.edu}}}
{mailto:garfinkl@oakland.edu}\\
WWW: \htmladdnormallink
{\protect {\tt{http://www.oakland.edu/?id=10223\&sid=249\#garfinkle}}}
{http://www.oakland.edu/?id=10223&sid=249\#garfinkle}\\

\section*{\noindent  Associate Editor\hfill}
Greg Comer\\
\smallskip
Department of Physics and Center for Fluids at All Scales,\\
St. Louis University,
St. Louis, MO 63103\\
Phone: (314) 977-8432\\
Internet:
\htmladdnormallink{\protect {\tt{comergl-at-slu.edu}}}
{mailto:comergl@slu.edu}\\
WWW: \htmladdnormallink{\protect {\tt{http://www.slu.edu/colleges/AS/physics/profs/comer.html}}}
{http://www.slu.edu//colleges/AS/physics/profs/comer.html}\\
\bigskip
\hfill ISSN: 1527-3431

\bigskip

DISCLAIMER: The opinions expressed in the articles of this newsletter represent
the views of the authors and are not necessarily the views of APS.
The articles in this newsletter are not peer reviewed.

\begin{rawhtml}
<P>
<BR><HR><P>
\end{rawhtml}
\end{flushleft}
\pagebreak
\section*{Editorial}

The next newsletter is due September 1st.  This and all subsequent
issues will be available on the web at
\htmladdnormallink 
{\protect {\tt {https://files.oakland.edu/users/garfinkl/web/mog/}}}
{https://files.oakland.edu/users/garfinkl/web/mog/} 
All issues before number {\bf 28} are available at
\htmladdnormallink {\protect {\tt {http://www.phys.lsu.edu/mog}}}
{http://www.phys.lsu.edu/mog}

Any ideas for topics
that should be covered by the newsletter, should be emailed to me, or 
Greg Comer, or
the relevant correspondent.  Any comments/questions/complaints
about the newsletter should be emailed to me.

A hardcopy of the newsletter is distributed free of charge to the
members of the APS Topical Group on Gravitation upon request (the
default distribution form is via the web) to the secretary of the
Topical Group.  It is considered a lack of etiquette to ask me to mail
you hard copies of the newsletter unless you have exhausted all your
resources to get your copy otherwise.

\hfill David Garfinkle 

\bigbreak

\vspace{-0.8cm}
\parskip=0pt
\section*{Correspondents of Matters of Gravity}
\begin{itemize}
\setlength{\itemsep}{-5pt}
\setlength{\parsep}{0pt}
\item John Friedman and Kip Thorne: Relativistic Astrophysics,
\item Bei-Lok Hu: Quantum Cosmology and Related Topics
\item Veronika Hubeny: String Theory
\item Beverly Berger: News from NSF
\item Luis Lehner: Numerical Relativity
\item Jim Isenberg: Mathematical Relativity
\item Lee Smolin: Quantum Gravity
\item Cliff Will: Confrontation of Theory with Experiment
\item Peter Bender: Space Experiments
\item Jens Gundlach: Laboratory Experiments
\item Warren Johnson: Resonant Mass Gravitational Wave Detectors
\item David Shoemaker: LIGO Project
\item Stan Whitcomb: Gravitational Wave detection
\item Peter Saulson and Jorge Pullin: former editors, correspondents at large.
\end{itemize}
\section*{Topical Group in Gravitation (GGR) Authorities}
Chair: Stan Whitcomb; Chair-Elect: 
Steve Detweiler; Vice-Chair: Patrick Brady. 
Secretary-Treasurer: Gabriela Gonzalez; Past Chair:  David Garfinkle;
Delegates:
Lee Lindblom, Eric Poisson,
Frans Pretorius, Larry Ford,
Scott Hughes, Bernard Whiting.
\parskip=10pt

\vfill
\eject

\section*{\centerline
{GGR program at the APS meeting in Washington D.C.}}
\addtocontents{toc}{\protect\medskip}
\addtocontents{toc}{\bf GGR News:}
\addcontentsline{toc}{subsubsection}{
\it GGR program at the APS meeting in Washington D.C., by David Garfinkle}
\parskip=3pt
\begin{center}
David Garfinkle, Oakland University
\htmladdnormallink{garfinkl-at-oakland.edu}
{mailto:garfinkl@oakland.edu}
\end{center}
We have an exciting GGR related program at the upcoming APS ``April'' meeting
(this year in February) in Washington D.C.  Our chair-elect
Steve Detweiler did an excellent job of putting together this program.  
At the APS meeting there will be several invited sessions of talks sponsored by the Topical Group in Gravitation (GGR).  
The large number of sessions sponsored by GGR means that our Topical Group 
has become one of the most important units at this meeting: only the Divisions
of Astrophysics, Particles and Fields, and Nuclear Physics have a larger 
presence at the April meeting.

The invited sessions sponsored by GGR are as follows:\\

Quantum Black Holes: Theory and Applications\\
Finn Larsen, Astrophysical Black Holes in String Theory?\\
Glenn Starkman, Realistic (?) Black Holes at the LHC\\
Sean Hartnoll, Holographic Approach to Condensed Matter Physics\\

Probing Strong-Field Gravity with Observations of the Galactic Center Black Hole\\
David Merritt, Probing Strong Field Gravity at the Galactic Center Using Stellar Motions\\
Sheperd Doeleman, Observing an Event Horizon: (sub)mm Wavelength VLBI of SgrA*\\
Avery Broderick, Nature of the Black Hole in the Center of the Milky Way\\

Numerical Relativity and Astrophysics\\
(joint with DCOMP)\\
Carlos Lousto, Statistical Studies of Spinning Black Hole Binaries\\
Matthew Duez, What Happens When Black Holes and Neutron Stars Merge?\\
Scott Noble, Seeing Spacetime by Proxy: Binary Black Holes in Gaseous Environments\\

Earth, Sky and Moon: Gravity Tests Across 13 Orders of Magnitude\\
(joint with GPMFC)\\
Stephan Schlamminger, Laboratory Tests of the Inverse Square Law of Gravity\\
Pierre Touboul, ESA's GOCE Gravity Gradiometer Mission\\
Tom Murphy, Advancing Tests of Relativity via Lunar Laser Ranging\\

Ground-based Interferometers on the Road to Gravitational Wave Astrophysics\\
Rana Adhikari, The LIGO and VIRGO Gravitational Wave Detectors\\
Peter Shawhan, Multi-Messenger Astronomy and Astrophysics with Grativational Wave Transients\\
Xavier Siemens, Science of Continuous Gravitational Wave Signals: Periodic Waves and the Stochastic Background\\
\vfill\eject

Gravity in Extreme Conditions\\
(joint with DCOMP)\\
Christian Ott, Computational Models of Stellar Collapse, Core-Collapse Supernovae, and Black Hole Formation\\
Frans Pretorius, Aneesur Rahman Prize for Computational Physics Talk: Black Hole Collisions\\
Steve Liebling, Status Report on Black Hole Critical Behavior\\

The GGR contributed sessions are as follows:

Lorentz Symmetry in Gravitation; Followed by LISA Developments\\
(joint with GPMFC)\\

Gravitational Collapse and Numerical Relativity\\

Observational Implications of Gravitational Waves\\
(joint with DAP)\\

Dynamics of Black Holes\\

Modeling Black Hole Binaries\\

Interpretation of Gravitational Wave Forms from Compact Binaries\\

Black Holes\\

Approximations in General Relativity\\

Numerical Simulations of Black Holes and Neutron Stars\\

Gravitational Waves from Neutron Stars\\

Quantum Aspects of Gravitation\\

Equivalence Principle and Precision Gravity Tests\\
(joint with GPMFC)\\

Advances in Ground-Based Gravitational Wave Detection\\

Foundational Aspects of General Relativity\\

\vfill\eject
\section*{\centerline
{we hear that \dots}}
\addtocontents{toc}{\protect\medskip}
\addcontentsline{toc}{subsubsection}{
\it we hear that \dots , by David Garfinkle}
\parskip=3pt
\begin{center}
David Garfinkle, Oakland University
\htmladdnormallink{garfinkl-at-oakland.edu}
{mailto:garfinkl@oakland.edu}
\end{center}

Frans Pretorius has been awarded the APS Aneesure Rahman Prize for Computational Physics.

Stefan Hollands has been awarded the Xanthopoulos Prize.

Nicolas Yunes has been awarded the Ehlers Thesis Prize.

Victor Taveras has been awarded the Bergmann-Wheeler Thesis Prize.

Manuella Campanelli was elected Vice-Chair of GGR; and Laura Cadonati and Luis Lehner were elected members at large of the GGR executive committee.

Manuela Campanelli, Karsten Danzmann, Katie Freese, Joseph Giaime, Jens Gundlach, 
Craig Hogan, Keith Riles, and Matt Visser have been elected as APS Fellows.

Hearty Congratulations!

\vfill\eject

\section*{\centerline
{Supernovae modelling}}
\addtocontents{toc}{\protect\medskip}
\addtocontents{toc}{\bf Research briefs:}
\addcontentsline{toc}{subsubsection}{
\it Supernovae modelling, by Ian Hawke}
\parskip=3pt
\begin{center}
Ian Hawke, University of Southampton
\htmladdnormallink{I.Hawke-at-soton.ac.uk}
{mailto:I.Hawke@soton.ac.uk}
\end{center}


Supernovae have repeatedly spurred scientific advances in
understanding the ``extra-solar'' universe. This was most recently
illustrated by SN1987A, the first supernova observed on Earth in every
electromagnetic band, as well as being the first object (except the
Sun) to be detected in neutrinos. The huge scientific effort to
understand these observations, reviewed near the time
by~\cite{Arnett:1990au,McCray:1993ga}, is ongoing with a range of
groups modelling the details of the collapse and explosion. With
suggestions that a galactic supernova could be detectable using
gravitational waves by current detectors such as LIGO, and that such a
detection may shed light on the supernova process (for a review
see~\cite{0264-9381-26-6-063001}), the specific importance of this
problem to the gravity community is clear; much of this article will
concentrate on this aspect. What is less clear remains the answer to
the central problem: what makes a supernova explode?


An outline of the early stages of collapse is simple enough to sketch
(for more detail see, e.g., \cite{Janka:2006fh}).  If a main sequence
star is sufficiently massive then eventually the pressure support from
nuclear burning is insufficient to support it against its own
self-gravity, leading to collapse. When nuclear densities ($\rho
\approx 10^{14} \, {\rm g} \, {\rm cm}^{-3}$) are reached the
degenerate matter is less compressible, leading to increased pressure
support in the core and an outwardly propagating ``bounce''
shock. This is driven from the edge of the decelerated core into the
outer layers which continue to fall supersonically in.

At this point the viability of the explosion becomes a competition
between the energy of the bounce shock and that of the infalling
matter. The strength of the shock is largely determined by the mass
and composition of the progenitor, but is also influenced by the
approximations in the modelling (e.g., the use of a relativistic
instead of a Newtonian gravity model leads to more compact cores;
interestingly, this does not necessarily lead to larger gravitational
wave amplitudes, as shown by~\cite{Dimmelmeier:2002bm}). Although some
early simulations with extreme progenitors or equations of state
suggested that a ``prompt'' shock explosion was possible, all current
calculations agree that the energy lost as the outgoing shock
dissociates the infalling matter to nucleons causes the bounce shock
to stall.

To initiate the explosion the shock needs to be revived. The precise
mechanism for this is uncertain, with different models and simulations
suggesting different key features. A central model is the neutrino
heating mechanism~(\cite{Colgate:1966ax,Bethe:1984ux}), where the
neutrinos remove energy from the hot core, some of which is deposited
behind the stalled shock, heating the matter and re-energizing the
shock. The existence of various hydrodynamical instabilities behind
the shock may amplify this mechanism away from spherical symmetry. A
more recent suggestion is the acoustic
mechanism~(\cite{Burrows:2005dv,Burrows:2006uh}) where strong g-modes
excited in the core during collapse transfer energy outwards via sound
waves, steepening into shocks and depositing their energy to revive
the stalled shock. Finally, there remain suggestions that magnetic
fields wound up during collapse could convert rotational to kinetic
energy leading to the explosion~(as reviewed by e.g.\
\cite{Kotake:2005zn}).

Modelling each of these scenarios and distinguishing between them
clearly requires detailed, high accuracy numerical simulations of
hydrodynamics in strong field gravity with radiation transport, taking
into account the composition and reactions of the stellar material and
also the magnetic field. The importance of convective and turbulent
motion also means a multi-dimensional treatment is essential. Despite
recent advances (illustrated by simulations such as
\cite{0004-637X-694-1-664,Murphy:2009dx,1742-6596-180-1-012018}) a
comprehensive treatment remains out of reach. With detailed tests (see
\cite{Marek:2005if,Mueller:2008it}) suggesting that a modified
Newtonian potential is a sufficient model of gravity to capture all of
the relevant features there is little need for full numerical
relativity. In addition, magnetic fields are not currently believed to
be an important factor in the explosion due to the slow rotation of
the best progenitor models (\cite{Heger:2004qp}). The modelling of the
neutrino transport and the core hydrodynamics is then key.

In recent simulations (see in particular~\cite{0004-637X-694-1-664})
the focus has moved away from the acoustic mechanism, which has been
difficult to reproduce away from the original scenarios, towards a
neutrino-driven explosion aided by convection (see
e.g.~\cite{Dessart:2005ck}) and dynamical instabilities behind the
stalled shock. The main feature is that the stalled accretion shock is
unstable to non-radial perturbations, with the low $\ell$ modes
showing the highest growth. This Standing Accretion Shock Instability
(SASI), first investigated by~\cite{Blondin:2002sm}, is now suggested
to be an essential driver of the explosion mechanism. The SASI causes
the shock to slosh back and forth, pushing the shock further out and
increasing the range over which neutrinos can deposit energy to revive
the shock. The precise nature of the interaction between the SASI and
the core remains unclear (see
e.g.~\cite{Foglizzo:2006fu,2006ApJ...642..401B}), as does the effects
of the progenitor mass and the importance of 3 rather than 2
dimensional numerical simulations (see \cite{1742-6596-180-1-012018}
for early results). What is clear is that the gravitational wave
signal post-bounce will contain considerably more information about
the physics (for results directly related to the SASI
see~\cite{Murphy:2009dx}) than the signal around bounce, which has
been shown by~\cite{Dimmelmeier:2008iq} to be fairly generic.


The achievement of multi-dimensional accurate numerical simulations of
core collapse with detailed approximations of radiation transport,
hydrodynamics and gravity have given hope that the essential
mechanisms behind supernova explosions are on the way to being
understood. The complexity and sheer length of time required for these
simulations are problems that can and will be overcome. Should we be
lucky enough for a repeat of SN1987A, the information gained through
combining gravitational wave detections with neutrinos and
electromagnetic signals could be spectacular.


\vfill\eject

\section*{\centerline
{ADM-50}}
\addtocontents{toc}{\protect\medskip}
\addtocontents{toc}{\bf Conference reports:}
\addcontentsline{toc}{subsubsection}{
\it ADM-50, 
by Richard Woodard}
\parskip=3pt
\begin{center}
Richard Woodard, University of Florida 
\htmladdnormallink{woodard-at-phys.ufl.edu}
{mailto:woodard@phys.ufl.edu}
\end{center}

2009 was the 50th anniversary of the development by Arnowitt, Deser 
and Misner of a canonical formalism for general relativity with 
asymptotically flat boundary conditions. This has been the basis 
for most subsequent work on the initial value problem in gravity, 
on numerical general relativity (as applied to neutron star and black
hole collisions), on the stability of general relativity and 
alternate gravity models, and even on cosmological perturbations. 
It was a major stimulation for the current experiments to detect 
gravitational waves, and has had many applications to the study of 
black holes in string theory. The importance of the ADM formalism was 
recognized by the American Physical Society through award of the 1994 
Dannie Heineman Prize in Mathematical Physics to Arnowitt, Deser and 
Misner.

A conference was held to mark the anniversary over the weekend of 
November 7-8. It took place in the newly completed Mitchell 
building on the campus of Texas A\&M University in College Station, 
Texas. Eighty-six participants attended two days of talks.
The slides of each speaker's talk are available on the conference
webpage at 
\htmladdnormallink
{\protect {\tt{http://adm-50.physics.tamu.edu}}}
{http://adm-50.physics.tamu.edu}\\
Just click on 
`Speakers' and the slides of each talk are there as a pdf file.

The opening talk was given by Jim Hartle, who spoke on work done in
the context of minisuperspace with Stephen Hawking and Thomas Hertog 
about identifying quantum gravitational states which predict 
approximately classical cosmologies. He closed with three questions
for ADM:
\begin{enumerate}
\item{Fifty years ago what did you think would be the future of
quantum gravity?}
\item{Where do we stand today and what do we have left to do?}
\item{How do you think today's efforts will appear at ADM-100?}
\end{enumerate}
Richard Woodard spoke next on the quantum generalization of the
remarkable demonstration by ADM that the self-energy of a classical, 
charged and gravitating particle is finite. In the subsequent question 
period Stanley Deser mentioned that this paper (Phys. Rev. Lett. 
{\bf 4} (1960) 375-377) was also the first work on general relativity 
ever published by Physical Review Letters! Jorge Pullin rounded out 
the Saturday morning session with report on work with Rodolfo Gambini 
and Rafael Porto on the problem of time in quantum gravity. He advocated 
a synthesis of the two main approaches: ``evolving Dirac observables'' 
and ``conditional probabilities.''

A scientific highlight of the conference came Saturday afternoon 
with back-to-back talks by Zvi Bern and Kelly Stelle on the recent 
demonstration that $N=8$ supergravity is on-shell finite at four loop 
order in $D=4$ and (the surprise) $D=5$. Bern described his heroic
computation with Carrasco, Dixon, Johansson and Roiban. While paying 
generous tribute to this work, Stelle explained how one can understand 
the cancellations, {\it ex post facto}, from field theoretic 
nonrenormalization theorems whose protection should fail at seven loop 
order in $D=4$ dimensions. Bern disagreed, and pointed to all orders 
results that can be obtained for a sub-class of diagrams.

Ben Gold from the WMAP collaboration presented the most recent 
results on the six cosmological parameters of the concordance 
model, and on the tensor-to-scalar ratio, the dark energy
equation of state, exotic perturbation models, running of the
scalar spectral index, neutrinos and non-Gaussianity.
Bob Wald concluded the Saturday session with a discussion of
the important auxiliary structure endowed upon classical field
theory by the existence of a Lagrangian or a Hamiltonian.

The conference banquet featured reminiscences of how ADM came to be.
Stanley Deser explained that he and Dick Arnowitt began working on
gravity because a spin two force carrier seemed like the next frontier
after the great successes of spin one QED. They talked of it at a
meeting at Neuchatel in the summer of 1958 where Pauli was present.
By that time they had reworked gravity into a 3+1 decomposition.
Then they were summoned to Princeton by John Wheeler who introduced
his ``brightest student since Feynman'' and Charlie Misner modestly
proposed ``some ideas that might be relevant.'' Misner took up the
story at this point, explaining how he met his Danish wife through
her sister having organized a meeting which he attended. The couple
were married in June of 1959 so they could go together to the island
of Bornholm where Stanley and his Swedish wife had decided to spend
the summer to avoid the heat of Copenhagen. The key ADM work was
done on Bornholm from June to August of 1959, although the trio
continued through 1961, partly at Brandeis, some at Syracuse and
much by phone and letters. At Bornholm they began work in an
elementary school which is still standing, though now as a tourist
office. When the Danish school started up again they transferred to
a now demolished Kindergarten where they had to crouch to use the tiny
blackboards. Some of their best work was done there.

Sunday began with Steve Carlip speaking on his suggestion that
any theory with approximate conformal invariance near the event 
horizon gives the Bekenstein-Hawking entropy. Next came Nick
Suntzeff who gave a wide ranging talk on using Type Ia Supernovae 
as tools for precision cosmology. He reminded us all of the history 
of the subject (``I am not part of the Harvard Group -- they are
a part of my group''), spoke on results from the ESSENCE Survey,
described the goals of the Carnegie Supernova Project and ended
by discussing modified gravity as an alternative to either dark
matter or dark energy. Rainer Weiss closed the morning session 
with a review of gravitational wave detection. Among many other things,
he noted that Stanley Deser sat on the 1983 NSF Advisory Committee 
which recommended LIGO, and was even one of three on the
Subcommittee which actually drafted the report.

A distinctive Texas flavor was added to Sunday's lunch by Phil 
Yasskin who organized an expedition to a local barbeque restaurant.
Gorged, and only a little late, we returned to hear Bernard Schutz 
describe numerical relativity analyses of compact mergers.
Although avoiding numerical instabilities requires reformulating the 
equations along the lines worked out by Baumgarte, Shapiro, Shibata, 
and Nakamura, the ``ADM notation and conceptualization remain the 
language of numerical relativity''. Chris Pope spoke on exact 
solutions to general relativity in higher dimensions. One point he
made, fortified by examples, is that some of the familiar 4-dimensional
uniqueness theorems no longer apply. Mike Duff closed the conference
by describing a correspondence between the entanglement measures of 
qubits in quantum information theory and black hole entropy in string
theory.

\end{document}